# All Electrical Control and Temperature Dependence of the Spin and Valley Hall Effect in Monolayer WSe$_2$ Transistors

AUTHOR NAMES

Xintong Li[†], Zhida Liu[‡], Yihan Liu[†,§], Suyogya Karki[†], Xiaoqin Li[‡], Deji Akinwande[†], and Jean Anne C. Incorvia[†]*

AUTHOR ADDRESS

[†]Department of Electrical and Computer Engineering, The University of Texas at Austin, Austin, Texas 78712, United States

[‡]Department of Physics and Center for Dynamics and Control of Materials, The University of Texas at Austin, Austin, Texas 78712, United States

*Corresponding Author, incorvia@austin.utexas.edu

KEYWORDS

Spin and valley Hall effect, Spatial Kerr rotation measurement, monolayer tungsten diselenide

ABSTRACT

Heavy metal-based two-dimensional van der Waals materials have a large, coupled spin and valley Hall effect (SVHE) that has potential use in spintronics and valleytronics. Optical measurements of the SVHE have largely been performed below 30 K and understanding of the SVHE-induced spin/valley polarizations that can be electrically generated is limited. Here, we study the SVHE in monolayer p-type tungsten diselenide (WSe$_2$). Kerr rotation (KR) measurements show the spatial distribution of the SVHE at different temperatures, its persistence up to 160 K, and that it can be electrically modulated via gate and drain bias. A spin/valley drift and diffusion model together with reflection spectra data is used to interpret the KR data and predict a lower-bound spin/valley lifetime of 4.1 ns below 90 K and 0.26 ns at 160 K. The excess spin and valley per unit length along the edge is calculated to be 109 $\mu m^{-1}$ at 45 K, which corresponds to a spin/valley polarization on the edge of 6%. These results are important steps towards practical use of the SVHE.

TEXT

The spin and valley degrees of freedom in two-dimensional (2D) van der Waals materials and heterostructures provide platforms for novel physics and potential applications in spintronics and valleytronics[1–7]. Comparing with graphene, transition metal dichalcogenide (TMDC) monolayers possess an intrinsically broken inversion symmetry, together with strong spin-orbit coupling (SOC), which leads to a coupled spin and valley Hall effect (SVHE) that makes the electrical generation and manipulation of spin and valley polarizations convenient[8–10]. Monolayer tungsten diselenide (WSe$_2$) features a significant valence band spin-splitting[11] (0.46 eV) and a $p$-type behavior in field effect transistors (FET), making it an ideal platform for electro-optical modulation of the spin and valley degrees of freedom[12]. The intervalley scattering of holes in WSe$_2$ must also flip spin, resulting in a prolonged spin and valley lifetime[13]. While all-electrical measurements of the spin or valley Hall effect have been performed at room temperature[14–16], in some of these methods it can be challenging to distinguish the SVHE non-local electrical signal from other potential sources. Optical measurements of electrically generated SVHE, on the other hand, can directly image the spatial distribution of the polarization and can be used to extract more parameters. Previous work has demonstrated, via spatial Kerr rotation (KR) measurements, the observation of the SVHE in n-type bilayer[17] and monolayer[18] MoS2 and in p-type monolayer WSe$_2$ FETs[19] at temperatures from 20-30 Kelvin. However, detailed interpretation of the SVHE-induced KR data and the study of the temperature and carrier density dependence of SVHE parameters is still lacking.

Here, we study the spatial distribution of the SVHE in monolayer WSe$_2$ FETs at different temperatures via spatial KR measurements, and show its persistence up to 160 K. We show the SVHE can be electrically modulated by gate and drain bias. By measuring the reflectance spectra vs. temperature, voltage, and probing laser energy, a conversion is developed between measured KR signal and spin/valley imbalance density. We then develop a spin/valley drift diffusion model to interpret the data and calculated the lower-bound of spin/valley lifetime, and estimate the spin/valley polarization on the edge of the channel. A more detailed comparison between this work and other work on SVHE can be found in supporting information S1.

In the WSe$_2$ transistor, a hole current flows through the WSe$_2$ channel when applying a drain bias and a negative gate bias to the transistor, and a spin/valley current is generated in the transverse direction by the SVHE, which causes spin/valley accumulation on the two edges that can be optically observed using linearly polarized light. The device structure and SVHE is illustrated in

Fig. 1a. Magneto-optical Kerr effect (MOKE) measurement is conducted to measure the spin/valley polarization distribution along a transverse line, without applying external magnetic fields. Figure 1b shows the schematic of the spin-split conduction band minima and valence band maxima near the K and K' valleys. The optical selection rule demands the absorption of circularly polarized light to be spin and valley dependent[9,20]. With incident linearly polarized light, a KR will occur in the reflected light if there is a local spin and valley imbalance in monolayer $WSe_2$, since the left- and right-circularly polarized components experience a different dielectric environment. Thus, the magnitude of the KR signal indicates spin/valley polarization, and the KR sign should be opposite for the two valleys, due to the opposite sign of the Berry curvature around K and K' valleys.

Figure 1c shows an optical image of one of the monolayer $WSe_2$ transistors. Fabrication process and images of more devices can be found in the Supporting Information S2. One of the typical spatial line scans of KR at 45 K is shown in Fig. 1d with line scan direction across the $WSe_2$ channel shown by the arrow in Fig. 1c. Opposite-sign peaks in KR are observed on opposite edges of the $WSe_2$ channel, showing the SVHE. The red line is a fitting curve that will be described below.

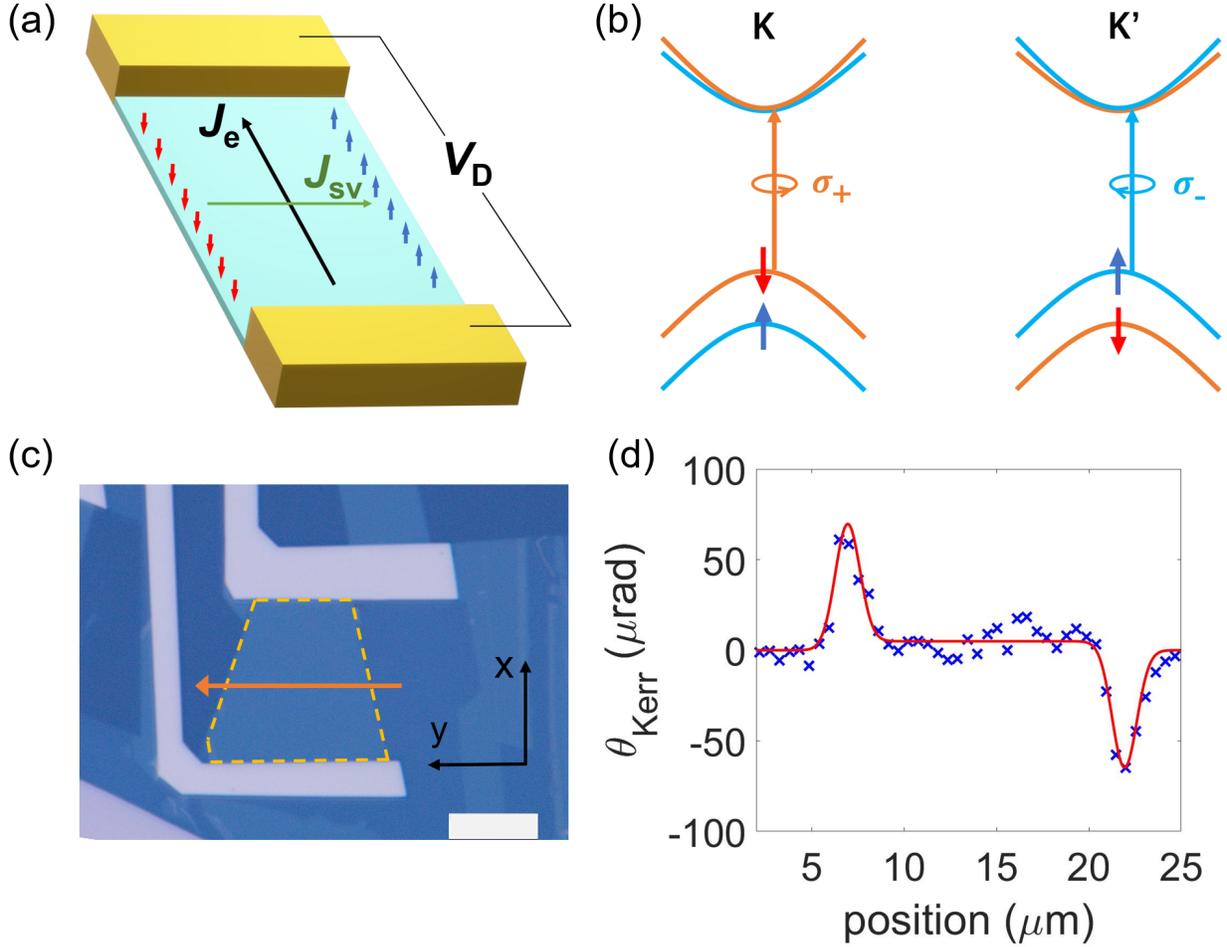

Figure 1. (a) Illustration of the device structure and SVHE. A source-drain voltage $V_D$ is applied on the source and drain contacts (yellow), with source grounded, and a negative gate voltage $V_G$ is applied to tune the doping level. A spin/valley current $J_{sv}$ is generated by SVHE which is perpendicular to the charge current $J_e$. (b) Schematic of the spin and valley dependence of the optical selection rule for left- ($\sigma_+$) and right- ($\sigma_-$) handed polarized lights. Arrows with colors mark the spin-split conduction band minima and valance band maxima at K and K' valleys. (c) Optical image of a typical WSe$_2$ FET device. Yellow dashed line marks the monolayer WSe$_2$ channel. Orange arrow marks the direction of the optical line scans. The white scale bar is 10 $\mu m$. (d) A typical spatial line scan of the KR measuring the SVHE with data (blue crosses), and fitting curve (red line) that accounts for the width of the Gaussian laser beam.

DC electrical characterization of monolayer WSe$_2$ FETs is conducted at different temperatures. Source-drain current $I_D$ vs. $V_G$ transfer characteristics, measured in ambient environment at room temperature (RT), are presented in Fig. 2a, for $V_D = 2$ V. A two-port field-effect hole mobility of $\mu = 0.17\ cm^2/(Vs)$ is extracted from the curve. This value is two orders of magnitude smaller than the phonon-scattering limited mobility at RT[21,22], mainly due to the detrimental effect of the contact resistance $R_C$ on the transconductance. Almost no $n$-type behavior is detected under a $V_G$ up to +20 V. Figure 2b shows the $I_D$ vs. $V_D$ characteristic of the same device at RT with various $V_G$ after intense KR measurement. Inset shows an $I_D$-$V_D$ curve measured at 45 K. The shape of the curve at low temperature indicates the existence of a Schottky barrier. Since the channel mobility should increase at lower temperatures in this range[23], we argue that $R_C$ becomes more dominant over the series resistance when cooling.

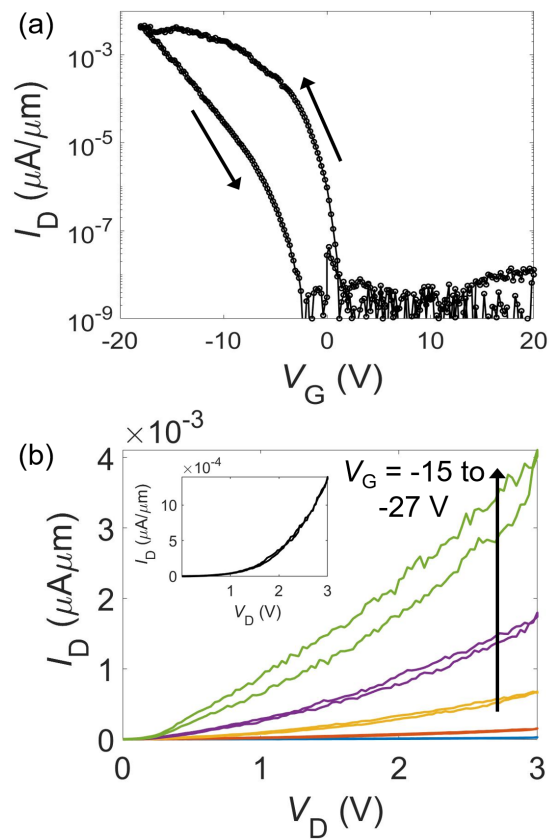

Figure 2. (a) Monolayer WSe$_2$ FET Room temperature (RT) $I_D$ vs. $V_G$ transfer characteristics measured in ambient environment. Clear $p$-type transport is observed while the $n$-type behavior is negligible for the given range of $V_G$. The hysteresis is common for unencapsulated TMDC transistors measured in air and was suppressed in the low-temperature, high-vacuum measurements. (b) $I_D$ vs. $V_D$ characteristic measured at RT with $V_G$ = -15, -18, -21, -24 and -27 V after intense KR measurements. Inset shows an $I_D$-$V_D$ curve measured at 45 K with $V_G$ = -22 V. The shape of the curve at low temperature indicates the existence of a Schottky barrier.

Gate voltage-dependent reflection spectra were measured on the monolayer WSe$_2$ transistor at selected temperatures with a white light source. Figure 3a shows a greyscale plot of the measured differential reflection spectra $\Delta R/R$ vs $V_G$ at 45 K, where $\Delta R/R = (R_{WSe_2} - R_{SiO_2})/R_{SiO_2}$ is the relative change of reflection of monolayer WSe$_2$ $(R_{WSe_2})$ with respect to the reflection of the Si/SiO$_2$ background $(R_{SiO_2})$. The peak at ~1.745 eV under small $V_G$ bias represents the absorption of neutral A exciton (X$^0$). When increasing (decreasing) $V_G$, a higher density of electrons (holes) is induced in the WSe$_2$ channel. X$^0$ tends to capture an additional electron (hole) and forms a negatively (positively) charged trion X$^-$ (X$^+$) that has smaller peak energy, shown by the reflection spectra peaks that appear at higher $V_G$ magnitudes. With increasing carrier density, the oscillator strength of X$^0$ decreases, while that of the trions increases. A broadening of the line width of X$^0$ absorption is also observed with increasing carrier density. More details can be found in Supporting information S4. We use a similar method from previous work[19,24] to fit these data and extrapolate the relation between the complex reflectivity and the carrier density of monolayer WSe$_2$. The complex dielectric function dispersion $\varepsilon(E) = \varepsilon_1(E) + i\varepsilon_2(E)$ is modeled as a superposition of Lorentzian oscillators:

$$\varepsilon(E) = 1 + \sum_k \frac{f_k}{E_k^2 - E^2 - iE\gamma_k} \qquad (1)$$

Here $f_k$, $E_k$ and $\gamma_k$ are the oscillator strength, resonant energy, and linewidth of the $k$th oscillator. In this treatment, instead of having $E_k$ run over the entire spectra, we only fit the oscillators corresponding to X$^0$ and X$^-$ (X$^+$) and write the rest of the terms as a constant $\varepsilon_b$. Using the standard transfer matrix analysis on thin film stacks[25], the first derivative of $\Delta R/R$ spectra is fit with the modeled function under measured $V_G$ and temperature. The WSe$_2$ monolayer is treated as a homogeneous medium with a thickness of 6.49 Å. Details of the modeling can be found in Supporting Information S5. Since the total hole density $p$ can be calculated from $V_G$ by $p = C_G(V_G - V_T)$, where $C_G$ is the gate oxide capacitance per unit area, and $V_T$ is the gate voltage at the charge neutral point, we obtained a relationship between $\varepsilon(E)$ and the total hole density $p$ at different temperatures.

The $\varepsilon(E)$ vs. $p$ relationship is used to predict how much KR angle ($\theta_{Kerr}$) can be generated from a given spin/valley sheet density imbalance $\Delta p_{sv}$, when a linearly polarized light impinges the WSe$_2$ sample under normal incidence. Details can be found in Supporting Information S6. For example, it is found that $\frac{\Delta p_{sv}}{\Delta \theta_{Kerr}} = 7.3 \times 10^7 \ cm^{-2}/\mu rad$ for small $\Delta p_{sv}$ under a $V_G$ of 22 V at 45 K, when probing with a 700 nm (1.771 eV) laser, which is close to the higher-energy side of X$^0$. This means that for a laser focused on an infinitesimal area of WSe$_2$, each $\mu rad$ of measured KR corresponds to a spin/valley sheet density imbalance of $7.3 \times 10^7 \ cm^{-2}$ at that area. This value is consistent with previous work[19]. We argue that the $\frac{\Delta p_{sv}}{\Delta \theta_{Kerr}}$ value is carrier density dependent (and therefore $V_G$ dependent) and should not be ignored in KR analysis. Figure 3b shows this modeled dependence by plotting the predicted KR angle with a spin/valley density imbalance of $1 \times 10^{10} \ cm^{-2}$ as a function of $V_G$, when probing with a 700 nm laser. This number $1 \times 10^{10} \ cm^{-2}$ is chosen as a typical value that yields a KR that is close to the measure value.

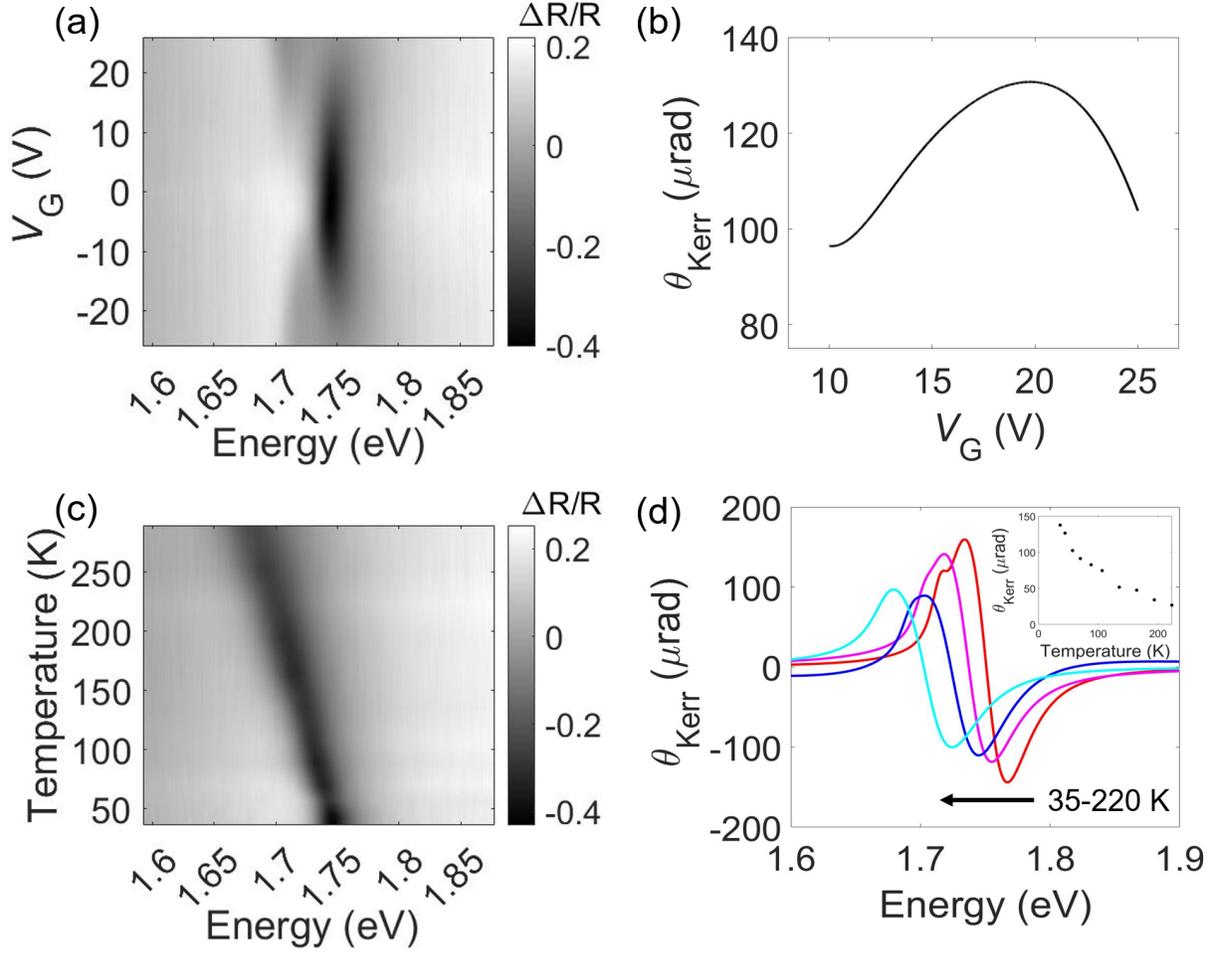

Figure 3. (a) Measured two-dimensional greyscale plot of the differential reflection spectra $\Delta R/R$ vs. $V_G$ at 45 K. The peak at ~ 1.745 eV under small $V_G$ bias shows the neutral A exciton ($X^0$). With increasing (decreasing) $V_G$, the $X^0$ peak gets weaker while the $X^-$ ($X^+$) peak with a smaller energy appears. (b) Modeled expected KR angle as a function of $V_G$, assuming a spin/valley density imbalance of $1 \times 10^{10}\ cm^{-2}$, with a 700 nm laser at 45 K. The Kerr rotation is expected to depend on $V_G$. (c) Measured differential reflection spectra $\Delta R/R$ as a function of temperature with $V_G = 0$ V. The $X^0$ peak redshifts from 1.745 eV at 45 K to 1.681 eV at 290 K. (d) Modeled expected KR angle with an assumed spin/valley density imbalance of $1 \times 10^{10}\ cm^{-2}$ as a function of photon energy at 35 K, 85 K, 130 K and 220 K. Inset shows the expected KR plotted against temperature for a photon energy of 1.745 eV.

The measured $\Delta R/R$ spectra as a function of temperature are shown in Fig. 3c with $V_G = 0$ V. The $X^0$ peak redshifts gradually from 1.745 eV at 45 K to 1.681 eV at 290 K, as the absorption peak gets broadened. The same modeling, as above, is conducted for every temperature measured. In Fig. 3d, the modeled expected KR angle with a spin/valley density imbalance of $1 \times 10^{10}\ cm^{-2}$ at $V_G = $ -22 V is plotted as a function of photon energy for a range of temperatures. The inset shows the expected KR probed by a 700 nm laser vs. temperature. Due to the redshift and broadening of the absorption peaks, the expected KR decreases with increasing temperature, if a laser with a wavelength of 700 nm is used. This effect must be carefully calibrated in temperature-dependent spin/valley polarization calculations. More details will be discussed in the following section.

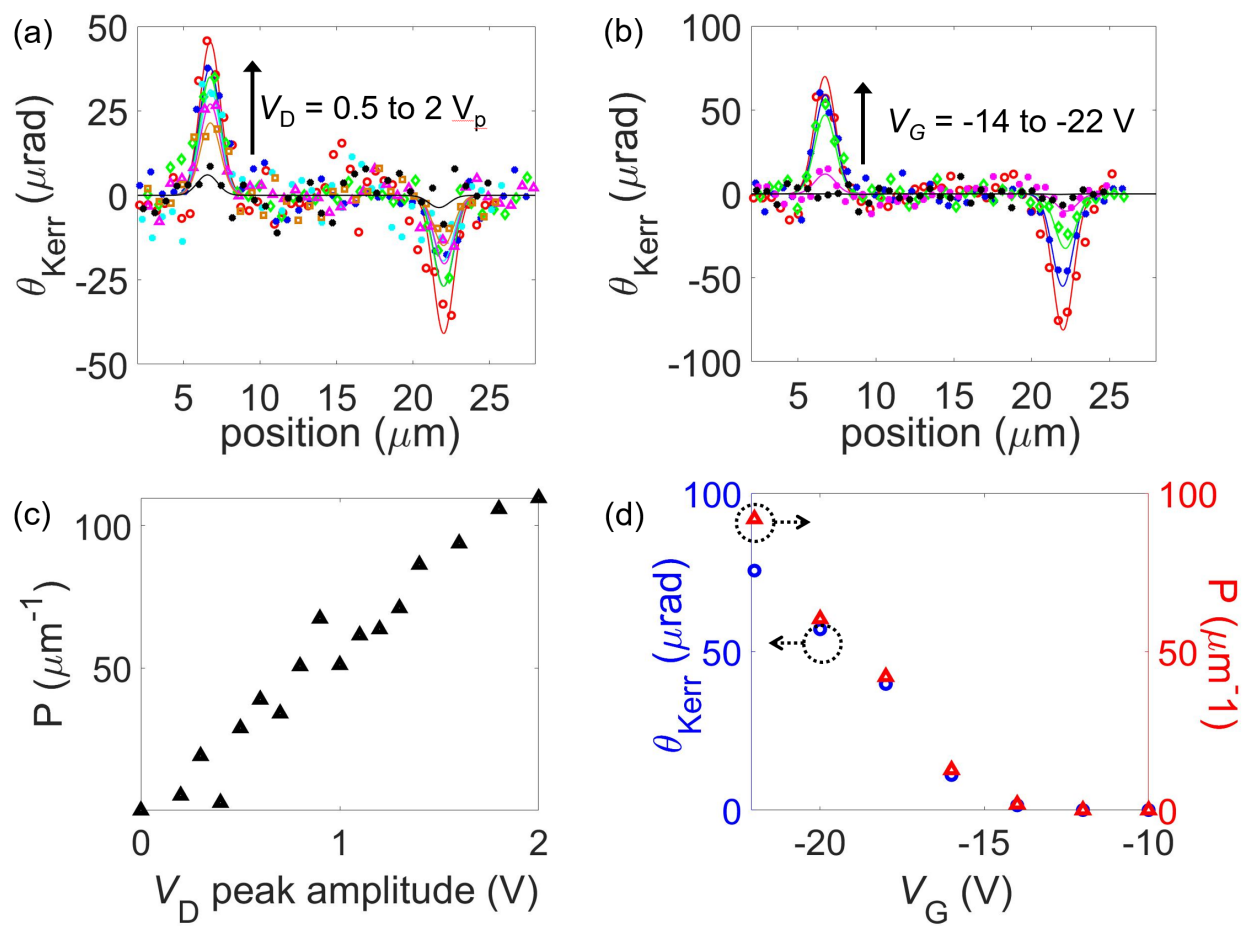

Figure 4. (a) Measured KR angle scanned along a transverse line on the WSe$_2$ channel as a function of $V_D$ at 100 K. $V_D$ peak voltage is chosen to be 0.5, 0.8, 1.1, 1.4, 1.6, 1.8 and 2.0 V while $V_G$ is held constant at -22 V. (b) KR angle scanned across the WSe$_2$ channel as a function of $V_G$ at 45 K. $V_G$ = -14 to -22 V with an interval of -2 V and $V_D$ is held to a constant AC amplitude 2 V$_p$. For (a) and (b) the markers represent measured data, and the lines are fitted curves considering the $fwhm$ of the laser. (c) Integrated spin/valley imbalance per unit length along the edge $P$, converted from the measured KR at the fixed spot on the edge where the KR is maximum, as a function of peak $V_D$. (d) Measured KR at the fixed spot on the edge where the KR is maximum, as a function of $V_G$ (blue circles) and corresponding calculated spin/valley imbalance near the edge per unit length $P$ (red triangles).

KR measurement of the SVHE-induced spin/valley polarization is performed in the monolayer WSe$_2$ transistor shown in Fig. 1c. Methods can be found in Supporting Information S4. No meaningful KR was detected on the SiO$_2$ background. Figures 4a-4b show the KR scanned along a line across the WSe$_2$ channel (direction depicted in Fig. 1c) as a function of $V_D$ and $V_G$, with fixed $V_G$ and $V_D$, respectively. The width of the WSe$_2$ channel is 16 $\mu m$ along this line. The marks are experimental data, and the curves are fits considering the finite full width half maximum ($fwhm$) of the Gaussian laser beam (details described below). The maximum of $V_G$ and peak $V_D$ are limited to -22 V and 2 V$_p$, respectively, to protect the device during measurement. We observe clear KR only on the WSe$_2$ edges, with opposite sign at opposite edges. Note that an overall negative KR offset across the sample is subtracted from the data. Examples of original data can be found in Supporting information S7. We argue that the offset is due to the magnetoelectric effect induced by the strain in the channel[18], and the drain voltage modulation of the carrier density in the channel, where an overall reflection modulation that is not fully canceled at the balanced photodetectors brings a finite KR reading. The consistency of the measured data and the fitted profile of the Gaussian laser beam indicate that the spin/valley accumulation at the edges spans only a small distance towards the middle. The KR at a fixed spot on the edge as a function of peak $V_D$ and $V_G$ measured at 45 K is present in Fig. 4c and 4d. This shows the SVHE can be controlled purely electrically for device applications.

We then use a spin/valley drift-diffusion model and recall the analysis in previous reflection measurements to interpret the measured KR at 45 K. In steady state, the spin/valley Hall current density $J_{sv}$ in the transverse direction is balanced by the spin/valley relaxation at the edges:

$$J_{sv} = \frac{ep_0 l_d}{\tau_{sv}} \quad (2)$$

where $e$ is the elementary charge, $p_0$ is the spin/valley density imbalance at the edges, $l_d$ is the diffusion length, and $\tau_{sv}$ is the spin/valley lifetime. Note that in the case of hole doping, we did not treat the spin Hall and valley Hall effect differently like what has been done in previous work, where spin Hall effect for electrons in $n$-doped MoS$_2$ are neglected due to the small spin splitting in the conduction band[17]. In $p$-doped WSe$_2$, due to its large spin-splitting in the valence band, both the spin Hall and valley Hall effects contribute greatly to the effective magnetization that generates the KR. So, in the following treatment we consider the spin and valley coupled ("spin/valley"). The spin/valley sheet density distribution along the transverse ($y$) direction can be modeled by

$$p_{sv}(y) = p_0 \left( e^{-\frac{y}{l_d}} - e^{\frac{y-w}{l_d}} \right) \tag{3}$$

Here $w$ is the channel width. In the measurements shown in Fig. 4a,b, we see that the KR is due to the spin/valley imbalance that is highly concentrated near the edge and $l_d$ cannot be detected by the spatial scan with finite laser $fwhm$, which suggests that $l_d \ll fwhm$. We can use equation (3) and the Gaussian distribution of the light intensity to convert the measured peak KR at the edges to $P \equiv p_0 l_d$, which is the integral of $p_{sv}(y)$ along the $y$ direction for either edge. Recall that in Fig. 3b,d, we provided a way to calculate the spin/valley sheet density imbalance $\Delta p_{sv}$ from the measured KR $\theta_{Kerr}$, and plotted the $V_G$ and temperature dependence of $\frac{\Delta p_{sv}}{\Delta \theta_{Kerr}}$. Then the integral $P$ can be calculated by

$$P = \sqrt{\frac{\pi}{4ln2}} fwhm \frac{\Delta p_{sv}}{\Delta \theta_{Kerr}} \theta_{Kerr} \tag{4}$$

under the condition of $l_d \ll fwhm$, where $\theta_{Kerr}$ is the KR angle measured directly in the MOKE test, while $\frac{\Delta p_{sv}}{\Delta \theta_{Kerr}}$ is the multiplier calculated from the reflection test. Because of the $V_G$ dependence of $\frac{\Delta p_{sv}}{\Delta \theta_{Kerr}}$ shown in Fig. 3b, it can be argued that the $P$ vs $V_G$ plot shown on the right side of Fig. 4d, instead of the $\theta_{Kerr}$ vs. $V_G$ plot shown on the left side, reflects the actual relationship between the spin/valley imbalance and $V_G$.

Combining the definition of $P \equiv p_0 l_d$ with equation (2),

$$P = \frac{\tau_{sv} J_{sv}}{e} = \frac{\tau_{sv} \sigma_{sv}}{e} E_x = \frac{\tau_{sv} \sigma_{sv}}{e} \frac{J_c}{\sigma_c} \tag{5}$$

Here $\sigma_{sv}$ is the spin/valley conductivity, $E_x$ is the electric field in the charge current direction ($x$), $J_c$ is the charge current density, and $\sigma_c$ is the charge conductivity of the channel. In these 2-port electrical measurements, the measured $\sigma_c$ is expected to be much lower than the actual $\sigma_c$ of the channel, and the calculated $E_x = V_d/L$ to be much larger than the actual $E_x$, due to the large contact resistance. Here $L$ is the channel length.

For the plot of $P$ vs. $V_D$ at 45 K shown in Fig. 4c, ideally there should be a linear relationship since $\tau_{sv}$ and $\sigma_{sv}$ depend little on $V_D$ bias, and $E_x$ is proportional to $V_D$ as shown in a previous work[18]. The Schottky barrier, however, should break the linear relationship between $E_x$ and $V_D$, especially at small bias, and make the $P - V_D$ curve more like the $I_D - V_D$ curve shown in Fig. 2b inset. The observed near-linear $P - V_D$ curve may rise from the difference of the AC $V_D$ used in the KR measurement and DC $V_D$ in the electrical test. We found that $P = 109 \, \mu m^{-1}$ at $V_D = 2 \, V$ and $V_G = -22 \, V$. Since the spin/valley mean free path $l_d$ cannot be accurately measured here, we use a reasonable $l_d = 100 \, nm$ from previous work[16], and use $p_0 = P/l_d$ to estimate the spin/valley density imbalance at the edge to be $1.09 \times 10^{11} \, cm^{-2}$. Gate voltage $V_G$ at the charge neutral point is estimated to be equal to the charge current threshold voltage $V_T = 10 \pm 3 \, V$; then, the total hole density $p = C_{ox}(V_G - V_T) = (1.73 \pm 0.52) \times 10^{12}\ cm^{-2}$. Thus the spin/valley polarization on the edge is approximately 6%.

The linearity between $P$ and $V_D$ can be used to estimate the lower bound of the spin (valley) lifetime $\tau_{sv}$, since we overestimated the actual electric field by $E_x = V_D/L$. The $\sigma_{sv}$ is predicted to be proportional to the hole density in the valence band when the doping level is small from a first principle study[26], and we estimated $\sigma_{sv} = 2.2 \times 10^{-4} e^2/\hbar$ for our $V_G$. With these parameters, the lower bound of the spin/valley lifetime is estimated to be 3.4 ns at 45 K. This value is near the lower end of previous work[13,27,28] as expected.

The $V_G$ dependence of $P$ at $V_D = 2\ V_p$ shown in Fig. 4d, is more complicated to interpret quantitatively. The linear relationship between $\sigma_{sv}$ and hole density is the dominant effect, which also explains that the $V_G$ at which KR starts to turn on is close to $V_T$. The non-linearity between $P$ and $V_G$ is due to the combined effect of $\tau_{sv}$ and $E_x$, where the $\tau_{sv}$ was reported to be negatively correlated to the carrier density but the dependence is relatively weak[29], and the actual $E_x = (V_D - I_D R_C)/L$ in the channel changes with $V_G$ due to the different response of the contact and channel resistance.

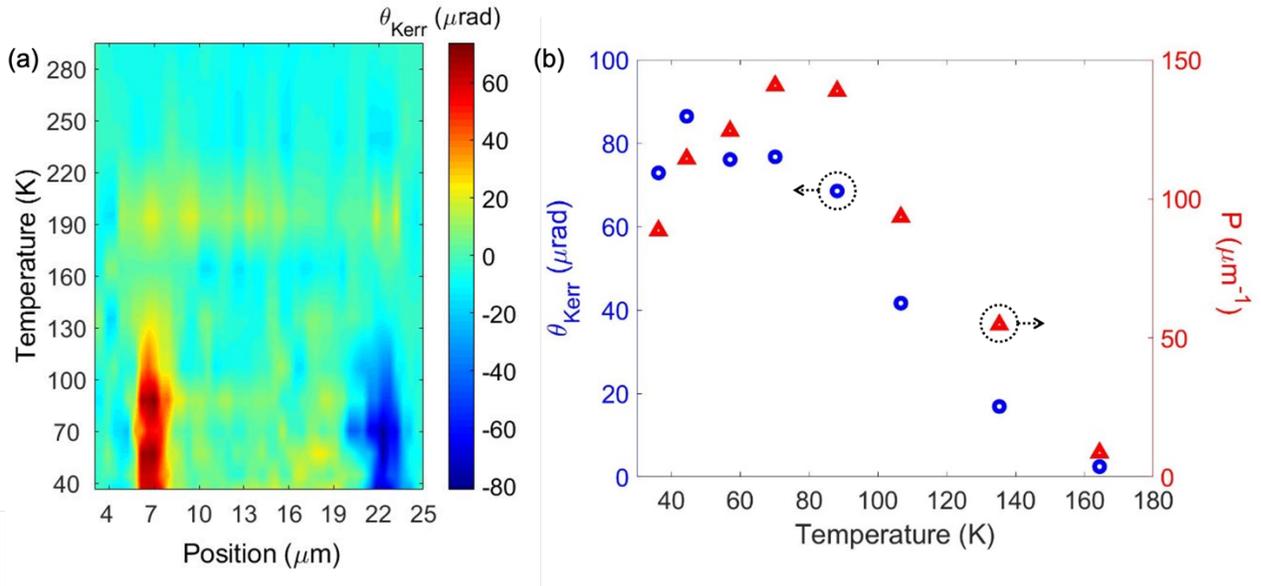

Figure 5. (a) KR spatial line scan as a function of temperature. Color represents magnitude of Kerr rotation in $\mu$rad. The KR peaks become undetectable above 160 K because the spin/valley polarization drops drastically, and the KR is buried in noise. (b) Measured average KR of the two peaks (blue circles) and corresponding calculated spin/valley imbalance per unit length $P$ (red triangles) vs. temperature.

We now study the evolution of the SVHE KR signal with increasing temperature. Figure 5a shows the KR spatial line scan as a function of temperature, probed by a 700 nm laser. The KR peaks shows opposite signs at opposite edges at all temperatures below 160 K. To our knowledge, this is the highest temperature to-date that the electrically-induced SVHE has been measured optically that shows a clear evidence of spin/valley accumulation at the edge. In the left axis of Fig. 5b, the average KR of the two peaks is plotted as a function of temperature, while on the right the measured KR is converted to the excess spin/valley number per unit length, $P$, accounting for the extraneous effects described throughout the text. We fully accounted for the $X^0$ peak redshift and broadening when increasing the temperature. We found that during cooling, $P$ appears at 160 K and grows fast until 90 K, where it starts to level and then decrease. For relatively high temperaures, $\tau_{sv}$ is dominated by phonon-assisted intervalley scattering[27] and drops fast with increasing temperature, while $\sigma_{sv}$ is reported to have relatively weak temperature dependence[30], resulting a fast decrease of $P$ from 90 K to 160 K. For the temperature below 90 K, we attribute the degradation of $P$ when lowering the temperature to the weaker temperature dependence of $\tau_{sv}$ combined with decreasing channel $E_x$. It was shown in previous work that the intervalley scattering at low temperatures is no longer dominated by phonon-assisted scattering, but is probably from scattering at grain boundaries and atomicaly sharp deformations[31,32], resulting in a weaker temperature dependence of $\tau_{sv}$. The temperature dependence of $E_x$ is explained in Supporting Information S8. For temperatures higher than 160 K, $\tau_{sv}$ and $P$ drop to a small value, and the resulting KR is buried in noise. We then roughly estimate $\tau_{sv}$ as a function of temperature by neglecting the weak temperature denpendence of $\sigma_{sv}$ and using an over-overestimated $E_x = V_d/L$, which results in $\tau_{sv} \approx 4.1$ ns at 90 K and $\tau_{sv} \approx 0.26$ ns at 160 K. Note that these are the lower bounds of the lifetime. For future KR analysis, we suggest a device structure for four-port resistance tests where the KR scan is conducted away from any contacts, so that mobility and $E_x$ can be extracted more accurately. Higher signal to noise is also required to obvserve the SVHE above 160 K.

In conclusion, we investigated the SVHE in monolayer WSe$_2$ transistors by measuring the spatial distribuition of the KR induced by the spin and valley imbalance. The KR data was converted to the excess spin and valley per unit length along the edge, $P$, by carefully fitting the reflection spectra at different temperatures and doping levels with a sum of the Lorentzian oscillator of $X^0$ and $X^+$. It was found that $P$ depends near-linearly on $V_D$ or $V_G$ bias, and decreases with increasing temperature above 90 K. We explained these results using a spin/valley drift and diffusion model. For example, at 45 K, it was measured $P = 109 \ \mu m^{-1}$, and if a spin/valley mean free path of 100 nm is assumed, this corresponds to a spin/valley polarization on the edge of 6%. A lower-bound spin or valley lifetime of 4.1 ns at 90 K and 0.26 ns at 160 K is estimated. We conclude that the manipulation of spin and valley degrees of freedom by SVHE at higher temperature is possible. Our findings demonstrate the potential of monolayer WSe$_2$ devices as a platform for electrically generated spin and valley polarization at different temperatures.

ASSOCIATED CONTENT

The supporting information is available free of charge on the ACS publications website.

Table of recent work on SVHE in few-layer TMD, fabrication process and images of more devices, electrical measurements, reflection spectra and KR measurement, modeling of the reflection spectra, converting KR angle to spin/valley accumulation P, raw data for KR measurements and temperature dependence of channel electrical field $E_x$ (PDF).

AUTHOR INFORMATION

**Corresponding Author**

*Email: incorvia@austin.utexas.edu**Present Addresses**

§Present Address: Department of Electrical and Computer Engineering, University of Pittsburgh, Pennsylvania 15261, United States

**Author Contributions**

The manuscript was written through contributions of all authors. All authors have given approval to the final version of the manuscript. X. L. fabricated the devices, carried out the measurements, and wrote the manuscript. Z. L. and Y. L. assisted with the measurements. S. K. contributed to the measurement setup. X.-Q. L. supervised and contributed to the measurement setup. D. A. supervised. J. A. I. conceived of the project, led supervising the work, and wrote the manuscript.

**Funding Sources**

This research was primarily supported by the National Science Foundation (NSF) through the Center for Dynamics and Control of Materials: an NSF MRSEC under Cooperative Agreement No. DMR-1720595. The optical measurement setup with supported in part by the NSF-Major Research Instrumentation Program (Grant MRI-2019130). This work was performed in part at the University of Texas Microelectronics Research Center, a member of the National Nanotechnology Coordinated Infrastructure (NNCI), which is supported by the National Science Foundation (Grant ECCS-2025227). The authors acknowledge the use of shared research facilities supported in part by the Texas Materials Institute and the Texas Nanofabrication Facility supported by NSF Grant No. NNCI-1542159.

**Notes**

The authors declare no competing financial interest.

ACKNOWLEDGMENT

We would like to thank Bin Fang and Tsun Chun Chang for their guide on the optical setup.